\documentclass[a4paper]{article}
\usepackage{graphicx}
\usepackage{amssymb}
\usepackage{amsmath}
\usepackage{bm}
\usepackage{layaureo}

\begin{document}

\title{The Brans-Dicke-Rastall theory}

\author{Thiago~R.~P.~Caram\^es,$^a$\footnote{tprcarames@gmail.com}~~Mahamadou~H.~Daouda,$^b$\footnote{daoudah8@yahoo.fr}~~J\'ulio~C.~Fabris,$^a$\footnote{fabris@pq.cnpq.br}\\ Adriano~M.~Oliveira,$^c$\footnote{adriano.ufes@gmail.com}~~Oliver~F.~Piattella,$^a$\footnote{oliver.piattella@pq.cnpq.br}~~and Vladimir~Strokov$^d$\footnote{vnstrokov@gmail.com}\vspace{0.5cm}\\
$a$ Departamento de F\'{\i}sica - UFES, Vit\'oria, ES, Brazil \\
$b$ D\'epartament de Physique - Universit\'e de Niamey, Niamey, Niger \\
$c$ IFES, Guarapari, ES, Brazil \\
$d$  Lebedev Physical Institute - Moscow, Russia,\\  \emph{Present address:} Departamento de F\'{\i}sica - UFES, Vit\'oria, ES, Brazil}

\maketitle

\begin{abstract}
We formulate a theory combining the principles of a scalar-tensor gravity and Rastall's proposal of a violation of the usual conservation laws. We obtain a scalar-tensor theory with two parameters $\omega$ and $\lambda$, the latter quantifying the violation of the usual conservation laws. The only exact spherically symmetric solution is that of Robinson-Bertotti  besides Schwarzschild solution. A PPN analysis reveals that General Relativity results
are reproduced when $\lambda = 0$. The cosmological case displays a possibility of deceleration/acceleration or acceleration/deceleration transitions during the matter dominated phase depending on the values of the free parameters.
\end{abstract}

\section{Introduction}

The Brans-Dicke theory~\cite{bd} appeared in the beginning of the sixties as an important alternative to the theory of General Relativity (GR). The main idea of this theory
is to consider the gravitational coupling $G$ as a dynamical quantity, implementing in this way the large number hypothesis formulated by Dirac~\cite{dirac}. Hence, a dynamical field $\phi$ represents the gravitational coupling, and it is introduced in the gravitational action through a kinetic term and a non-minimal coupling with the usual Ricci scalar.
A new parameter $\omega$ quantifies the interaction of the scalar field and the gravitational term, such that as \mbox{$\omega \rightarrow \infty$} the General Relativity
theory is recovered. The observational estimations obtained indicate a very large value for $\omega$, making the Brans-Dicke theory, in practice, very similar
to GR. Recent estimates using the PLANCK data point to a value $\omega \sim 1000$~\cite{skordis}. Local tests based on the PPN approach may lead to higher values of $\omega$~\cite{will}.
\par
In spite of those observational constraints, small -- or even negative -- values of the parameter $\omega$ may be very interesting. First of all, they sometimes arise in string theories in their low-energy limit \cite{cordas}. When negative values of $\omega$ are allowed, primordial singularity-free solutions emerge naturally from Brans-Dicke theory. Late time accelerated solution can be achieved \cite{f1,f2}, but at the price of a negative gravitational coupling. This last feature limits, of course, the attraction of such scenarios.
\par
We have recently been interested in some generalizations of GR that evoke the gravitational anomaly effect, viz. Rastall's theory~\cite{rastall1,rastall2}. These generalisations touch one of the cornerstones of gravity theories: the conservation laws encoded in the null divergence of the energy-momentum tensor. Since the concept of energy in GR is an object of discussion~\cite{energia,energia-err}, the possibility that the energy-momentum tensor has a non-zero divergence should be considered in some situations. For example, the chameleon mechanism \cite{camaleao} uses this possibility by re-expressing a scalar-tensor theory (like the Brans-Dicke one), originally formulated in Jordan's frame, in Einstein's frame. Also, quantum effects in a curved space-time may lead to a violation of the classical conservation laws \cite{quantum}.
\par
Rastall's theory leads to many interesting results when applied, for example, to the present universe \cite{grupo}. In a way, this theory can be viewed as a natural implementation of an interaction model in the dark sector of the present stage of the cosmic evolution. Alternatively, it can be considered as a mechanism to generate effective equations
of state when ordinary fields are considered in a curved space-time \cite{kerner}.
\par
Smalley~\cite{smalley} addressed the idea of violation of the conventional conservation laws in the context of Brans-Dicke theory. In this approach, the Klein-Gordon type equation for the scalar field was kept fixed while the Einstein equation changed accordingly. Here, we would like to revisit this proposal following a different path: we try to write down the field equations in such a way that Brans-Dicke and GR as well as the ordinary Rastall theory are recovered. The final equations seem to be simpler than those of reference \cite{smalley}.
\par
In this work we study this Brans-Dicke-Rastall (BDR) theory. We investigate the resulting field equations in two situations: spherically symmetrical and cosmological configurations. In the former case, we obtain that the only non-trivial solution is represented by the Robinson-Bertotti metric (its interpretation, however, differs from the conventional one). A solution that represents a star-like configuration is the ``trivial'' Schwarzschild one. A PPN analysis shows a
possibility of agreement with the usual tests of gravity theories. For some cases, the General Relativity results are reproduced. At cosmological level, we show that accelerated solutions are possible in the dust phase of the cosmic evolution without introducing dark energy. We display examples where a decelerated/accelerated or accelerate/decelerated transitions are achieved with a positive effective gravitational coupling.
\par
This paper is organised as follows. In the next section, we set up the field equations of the BDR theory. In Section~\ref{spherical} we analyze static spherically symmetric solutions while Section~\ref{PPN} covers the PPN analysis. In Section~\ref{cosmology} the cosmological context is addressed. Finally, in Section~\ref{conclusions} we summarize our conclusions.

\section{The theory}\label{theory}

The main idea of Rastall's theory \cite{rastall1,rastall2} is the assumption that in curved space-time the usual conservation laws used in GR are violated.
Hence, there must be a connection between the divergence of the energy-momentum tensor and the curvature of the space-time.
According to this program, the divergence of the energy-momentum tensor may be written as
\begin{eqnarray}
\label{r1a}
{T^{\mu\nu}}_{;\mu} = \frac{1 - \lambda}{16\pi G}R^{,\nu}.
\end{eqnarray}
In equation (\ref{r1a}) $\lambda$ is a free parameter codifying the deviation from the conservation. When $\lambda = 1$ the traditional conservation laws are recovered. Equation (\ref{r1a}) is a phenomenological way to implement the gravitational anomaly due to quantum effects (see~\cite{anomalia}, for example).
\par
In the context of the Brans-Dicke theory, we can make the identification:
\begin{equation}
G = \frac{1}{\phi}.
\end{equation}
Hence,
\begin{eqnarray}
{T^{\mu\nu}}_{;\mu} = \frac{(1 - \lambda)\phi}{16\pi}R^{,\nu}.
\end{eqnarray}
Let us generalize Rastall's version of the field equations to the Brans-Dicke case. Following the original formulation in the context of GR, a minimal modification implies:
\begin{eqnarray}
\label{fe}
R_{\mu\nu} - \frac{\lambda}{2}g_{\mu\nu}R = \frac{8\pi}{\phi}T_{\mu\nu} + \frac{\omega}{\phi^2}\biggr\{\phi_{;\mu}\phi_{;\nu} - \frac{1}{2}g_{\mu\nu}\phi_{;\rho}\phi^{;\rho}\biggl\}
+ \frac{1}{\phi}(\phi_{;\mu;\nu} - g_{\mu\nu}\Box\phi).
\end{eqnarray}
It is important to remark that even if the structure of the right hand side is the same as in the Brans-Dicke theory, the whole equation (\ref{fe}) can be derived from a Lagrangian only when $\lambda=1$.
\par
The trace of these ``Einsteinian equations'' reads:
\begin{eqnarray}
\label{escalar}
R = \frac{1}{1 - 2\lambda}\biggr\{\frac{8\pi}{\phi}T - \frac{\omega}{\phi^2}\phi_{;\rho}\phi^{;\rho} - 3\frac{\Box\phi}{\phi}\biggl\}.
\end{eqnarray}
With the aid of this expression equation (\ref{fe}) can be rewritten as
\begin{eqnarray}
R_{\mu\nu} - \frac{1}{2}g_{\mu\nu}R &=& \frac{8\pi}{\phi}\biggr\{T_{\mu\nu} - \frac{1 - \lambda}{2(1 - 2\lambda)}g_{\mu\nu}T\biggl\}+ \nonumber\\
&+& \frac{\omega}{\phi^2}\biggr\{\phi_{;\mu}\phi_{;\nu} + \frac{\lambda}{2(1 - 2\lambda)}g_{\mu\nu}\phi_{;\rho}\phi^{;\rho}\biggl\} + \nonumber \\
&+&\frac{1}{\phi}\biggr\{\phi_{;\mu;\nu} +
\frac{(1 + \lambda)}{2(1 - 2\lambda)}g_{\mu\nu}\Box\phi\biggl\}.
\end{eqnarray}
The Bianchi identities lead to
\begin{eqnarray}
\Box\phi = \frac{8\pi\lambda}{3\lambda - 2(1 - 2\lambda)\omega}T - \frac{\omega(1 - \lambda)}{3\lambda - 2(1 - 2\lambda)\omega}\frac{\phi^{;\rho}\phi_{;\rho}}{\phi}.
\end{eqnarray}
\par
The complete set of equations is:
\begin{eqnarray}
{T^{\mu\nu}}_{;\mu} &=& \frac{(1 - \lambda)\phi}{16\pi}R^{,\nu},\\
R_{\mu\nu} - \frac{1}{2}g_{\mu\nu}R &=& \frac{8\pi}{\phi}\biggr\{T_{\mu\nu} - \frac{1 - \lambda}{2(1 - 2\lambda)}g_{\mu\nu}T\biggl\} +\nonumber\\
&+& \frac{\omega}{\phi^2}\biggr\{\phi_{;\mu}\phi_{;\nu} + \frac{\lambda}{2(1 - 2\lambda)}g_{\mu\nu}\phi_{;\rho}\phi^{;\rho}\biggl\} + \nonumber \\
&+&\frac{1}{\phi}\biggr\{\phi_{;\mu;\nu} +
\frac{(1 + \lambda)}{2(1 - 2\lambda)}g_{\mu\nu}\Box\phi\biggl\},\\
\Box\phi &=& \frac{8\pi\lambda}{3\lambda - 2(1 - 2\lambda)\omega}T - \frac{\omega(1 - \lambda)}{3\lambda - 2(1 - 2\lambda)\omega}\frac{\phi^{;\rho}\phi_{;\rho}}{\phi}.
\end{eqnarray}
When $\lambda = 1$, the usual Brans-Dicke theory is recovered.
\par
Following the same steps as in the determination of the effective gravitational coupling today in~\cite{will,weinberg}, we find the following expression (see also Section~\ref{PPN}):
\begin{eqnarray}
\label{efetivo}
G = \frac{2[2\lambda + (3\lambda - 2)\omega)]}{3\lambda - 2(1 - 2\lambda)\omega}\frac{1}{\phi}.
\end{eqnarray}
When $\lambda = 1$ we obtain the corresponding expression for the Brans-Dicke theory.

\section{Spherically symmetric static vacuum solutions}\label{spherical}

The classical tests of theory of gravity are based on the motion of test particles in the geometry of a spherically symmetric object like a star or a planet.
Hence, to verify the viability of the theory proposed, it is crucial to look for a spherically symmetric solution. As a first step, the vacuum solution representing
the space-time in the exterior of a star-like object is considered.
\par
In the vacuum case, the equations reduce to
\begin{eqnarray}
R^{,\nu} &=& 0,\\
R_{\mu\nu} - \frac{1}{2}g_{\mu\nu}R &=&
\frac{\omega}{\phi^2}\biggr\{\phi_{;\mu}\phi_{;\nu} + \frac{\lambda}{2(1 - 2\lambda)}g_{\mu\nu}\phi_{;\rho}\phi^{;\rho}\biggl\} + \nonumber \\
&+&\frac{1}{\phi}\biggr\{\phi_{;\mu;\nu} +
\frac{(1 + \lambda)}{2(1 - 2\lambda)}g_{\mu\nu}\Box\phi\biggl\},\\
\label{v-e3}
\Box\phi &=& - \frac{\omega(1 - \lambda)}{3\lambda - 2(1 - 2\lambda)\omega}\frac{\phi^{;\rho}\phi_{;\rho}}{\phi}.
\end{eqnarray}
The first of these equations leads to,
\begin{equation}
R = R_0 = \mbox{constant}.
\end{equation}
Hence, in vacuum the Ricci scalar is necessarily constant. The case $R_0 = 0$ corresponds to the Schwarzschild solution of GR.

\subsection{Equations of motion}

Let us consider a metric in the form:
\begin{eqnarray}
ds^2 = e^{2\gamma}dt^2 - e^{2\alpha}dr^2 - e^{2\beta}(d\theta^2 + \sin^2\theta d\phi^2).
\end{eqnarray}
The functions $\alpha$, $\beta$ and $\gamma$ depend on the radial coordinate $r$ only.
Under this assumption, the only non-zero components of the Christoffel symbols are the following:
\begin{eqnarray}
\Gamma^0_{0r} &=& \gamma', \quad \Gamma^r_{00} = e^{2(\gamma - \alpha)}\gamma', \\
\Gamma^r_{rr} &=& \alpha', \quad \Gamma ^r_{\theta\theta} = - e^{2(\beta - \alpha)}\beta', \\
\Gamma^r_{\phi\phi} &=& - e^{2(\beta - \alpha)}\beta'\sin^2\theta, \quad \Gamma^\theta_{\phi\phi} = - \sin\theta\cos\theta,\\
\Gamma^\theta_{r\theta} &=& \Gamma^\phi_{r\phi} = \beta', \quad \Gamma^\phi_{\theta\phi} = \cot\theta.
\end{eqnarray}
\par
The non-zero components of the Ricci tensor are:\,\footnote{We use the definitions:
$R_{\mu\nu} = \partial_\rho\Gamma^\rho_{\mu\nu} - \partial_\nu\Gamma^\rho_{\mu\rho} + \Gamma^\rho_{\mu\nu}\Gamma^\sigma_{\rho\sigma} - \Gamma^\rho_{\mu\sigma}\Gamma^\sigma_{\rho\mu}$, $R = g^{\mu\nu}R_{\mu\nu}$\,, and $G^{\mu}_{\nu}=R^{\mu}_{\nu}-\frac{1}{2}\delta^{\mu}_{\nu}R$.}
\begin{eqnarray}
R_{0}^{0} &=& e^{- 2\alpha}[\gamma'' + \gamma'(\gamma' - \alpha' + 2\beta')],\\
R_{r}^{r} &=& e^{- 2\alpha}[\gamma'' + 2\beta'' - \alpha'(\gamma' + 2\beta') + \gamma'^2 + 2\beta'^2],\\
R_{\theta}^{\theta} &=& R_{\phi}^{\phi}=e^{- 2\alpha}[\beta'' + \beta'(\gamma' - \alpha' + 2\beta')] - e^{-2\beta}\,.
\end{eqnarray}
With these expressions, we can determine the Ricci scalar:
\begin{eqnarray}
R = e^{-2\alpha}[2\gamma'' + 4\beta'' + 2\gamma'(\gamma' - \alpha' + 2\beta') + 6\beta'^2 - 4\beta'\alpha'] - 2e^{-2\beta}.
\end{eqnarray}
\par
Now, we can determine the components of the Einstein:
\begin{eqnarray}
G_{0}^{0} &=& e^{-2\alpha}[- 2\beta'' + 2\alpha'\beta' - 3\beta'^2] + e^{- 2\beta},\\
G_{r}^{r} &=& -e^{-2\alpha}[2\beta'\gamma' + \beta'^2] + e^{ - 2\beta},\\
G_{\theta}^{\theta} &=& G_{\phi}^{\phi}=-e^{ - 2\alpha}[\gamma'' + \beta'' + \beta'(\gamma' - \alpha' + \beta') + \gamma'(\gamma' - \alpha')].
\end{eqnarray}
\par
Combining (\ref{escalar}) with (\ref{v-e3}) we find:
\begin{equation}
\label{curva}
R_0 = \omega\biggr\{\frac{3 + 2\omega}{3\lambda - 2(1 - 2\lambda)\omega}\biggl\}\frac{\phi_{;\rho}\phi^{;\rho}}{\phi^2}.
\end{equation}
In this expression we have used explicitly $R = R_0 =$ constant.
\par
Now, the D'Alambertian reads:
\begin{equation}
\label{box1}
\Box\phi = \frac{\left(\sqrt{-g}g^{\mu\nu}\phi_{,\nu}\right)_{,\mu}}{\sqrt{-g}}= - e^{-2\alpha}[\phi'' + (\gamma' + 2\beta' - \alpha')\phi']\,,
\end{equation}
where $g\equiv\det{g_{\mu\nu}}$.

\subsection{Integrating the equations of motion}

Let us choose the radial coordinate such that,
\begin{equation}
\alpha = \gamma + 2\beta.
\end{equation}

With this choice, the components of the Ricci tensor and the D'Alambertian simplify to
\begin{eqnarray}
R_{00} &=& e^{-4\beta}\gamma'', \\
R_{rr} &=& - \gamma'' - 2\beta'' + 4\beta'\gamma' + 2\beta'^2,\\
R_{\theta\theta} &=&R_{\phi\phi}/\sin^{2}{\theta}= - e^{-2(\gamma + \beta)}\beta'' + 1, \\
\Box\phi & =& - e^{-2\alpha}\phi''.
\end{eqnarray}

Let us write the Einsteinian equations as
\begin{align}
\label{eeq}
R_{\mu\nu} = \frac{\omega}{\phi^2}\biggr\{\phi_{;\mu}\phi_{;\nu} + \frac{\lambda - 1}{2(1 - 2\lambda)}g_{\mu\nu}\phi_{;\rho}\phi^{;\rho}\biggl\}
+ \frac{1}{\phi}\biggr\{\phi_{;\mu;\nu} + \frac{\lambda - 2}{2(1 - 2\lambda)}g_{\mu\nu}\Box\phi\biggl\}\,
\end{align}
or, in the extended form:
\begin{align}
\label{fe-1}
\gamma'' + \frac{\phi'}{\phi}\gamma' &= - \omega\frac{(\lambda - 1)}{2(1 - 2\lambda)}\biggr(\frac{\phi'}{\phi}\biggl)^2 - \frac{\lambda - 2}{2(1 - 2\lambda)}\frac{\phi''}{\phi},  \\
\label{fe-2}
\gamma'' + 2\beta'' - 2\beta'(\beta' + 2\gamma') - \frac{\phi'}{\phi} (\gamma' + 2\beta') &= - \omega\frac{1 - 3\lambda}{2(1 - 2\lambda)}\biggr(\frac{\phi'}{\phi}\biggl)^2 + \frac{3\lambda}{2(1 - 2\lambda)}\frac{\phi''}{\phi},\\
\label{fe-4}
\beta'' + \beta'\frac{\phi'}{\phi} - e^{2(\gamma + \beta)} &= -\omega\frac{\lambda - 1}{2(1 - 2\lambda)}\biggr(\frac{\phi'}{\phi}\biggl)^2 - \frac{\lambda - 2}{2(1 - 2\lambda)}\frac{\phi''}{\phi}.
\end{align}

Equations (\ref{v-e3}) and (\ref{curva}) lead to two supplementary equations:
\begin{eqnarray}
\label{r0}
R_0 &=& - \omega\biggr\{\frac{3 + 2\omega}{3\lambda - 2(1 - 2\lambda)\omega}\biggl\}e^{-2\alpha}\biggr(\frac{\phi'}{\phi}\biggl)^2,\\
\phi'' &=& - \omega\frac{1 - \lambda}{3\lambda - 2(1 - 2\lambda)\omega}\frac{\phi'^2}{\phi}.
\end{eqnarray}
Their solution is
\begin{eqnarray}
\label{s-1}
\phi &=& \phi_{0} (r/r_0)^\frac{1}{1 - A},\\
\alpha &=& - \ln{(r/r_0)} - \ln\biggr[(1 - A)\sqrt{\frac{R_0(1 - \lambda)}{(3 + 2\omega)A}}\biggl],
\end{eqnarray}
with $\phi_0$ and $r_0$ being integration constants.\,\footnote{Note that we have chosen $r=0$ as a reference point. Depending on the sign of $(1-A)$ it corresponds either to $\phi=0$ (infinitely strong gravity) or $\phi=\infty$ (no gravity).} We have also defined
\begin{equation}
\label{A}
A = - \omega\frac{1 - \lambda}{3\lambda - 2(1 - 2\lambda)\omega}.
\end{equation}

Using solution (\ref{s-1}) in equation (\ref{fe-1}), we find for $\gamma$ the following expression:
\begin{equation}
\gamma = \gamma_0 + \gamma_{1}(r/r_0)^{-\frac{ A}{1 - A}} + \gamma_2\ln{(r/r_0)},
\end{equation}
where $\gamma_0$ and $\gamma_1$ are arbitrary constants and
\begin{equation}
\label{gamma2}
\gamma_2 = - \frac{1}{2(1 - 2\lambda)}\biggr\{\frac{\omega(\lambda - 1) + (\lambda - 2)A}{A(1 - A)}\biggl\}.
\end{equation}

Now, using that $\beta = \frac{1}{2}(\alpha - \gamma)$, we obtain:
\begin{eqnarray}
\beta &=& -\frac{1}{2}\biggr\{ \ln{(r/r_0)} + \ln\biggr[(1 - A)\sqrt{\frac{R_0(1 - \lambda)}{(3 + 2\omega)A}}\biggl] + \gamma_0 + \gamma_{1}(r/r_0)^{-\frac{ A}{1 - A}} - \gamma_2\ln{(r/r_0)}\biggl\},\nonumber\\
&=& - \frac{1}{2}\biggr\{ \ln\biggr[(1 - A)\sqrt{\frac{R_0(1 - \lambda)}{(3 + 2\omega)A}}\biggl] + \gamma_0 + \gamma_{1}r^\frac{- A}{1 - A} + (\gamma_2 + 1)\ln{(r/r_0)}\biggl\},\nonumber \\
&=& \frac{1}{2}\biggr\{\alpha_0 - \gamma_0 - \gamma_{1}(r/r_0)^{-\frac{A}{1 - A}} - (\gamma_2 + 1)\ln{(r/r_0)}\biggl\},
\end{eqnarray}
where
\begin{equation}
\alpha_0 =  - \ln\biggr[(1 - A)\sqrt{\frac{R_0(1 - \lambda)}{(3 + 2\omega)A}}\biggl],
\end{equation}
is a constant.

Substituting the results in equation (\ref{fe-4}), we see that it is satisfied for any $r$ only if  $\gamma_1 = 0$ and $\gamma_2 = - 1$. This implies that the metric function $\beta$ is, in fact, constant. For the other metric functions the relations above yield:
\begin{eqnarray}
\label{alpha_beta_gamma}
\alpha &=& \alpha_0 - \ln{(r/r_0)},\\
\gamma &=& \gamma_0 - \ln{(r/r_0)},\\
\beta &=& \beta_0 = \frac{1}{2}(\alpha_0 - \gamma_0).
\end{eqnarray}
Hence, the metric is
\begin{equation}
ds^2 = e^{2\alpha_0}\frac{dt^2}{(r/r_0)^2} - e^{2\gamma_0}\frac{dr^2}{(r/r_0)^2} - e^{\alpha_{0}-\gamma_{0}}d\Omega^2.
\end{equation}
If the scale~$r_0$ is chosen such that $r_{0}^{2}=e^{\alpha_{0}-3\gamma_{0}}$, making redefinitions $t\rightarrow e^{-(\gamma_{0}+\alpha_{0})/2}$, $s\rightarrow e^{-(\gamma_{0}-\alpha_{0})/2}$, and $r\rightarrow rr_{0}$, we arrive at:
\begin{equation}
\label{sol-1}
ds^2 = \frac{1}{r^{2}}\left(dt^2 - dr^2 - r^{2}d\Omega^2\right)\,,
\end{equation}
which is the so-called Robinson-Bertotti solution~\cite{robinson,bertotti} that is obtained, in the context of GR, by considering an electromagnetic field. Hence, no black hole solution is possible. This solution appears as the only non-trivial (non-Schwarzschild solution) vacuum solution.

Interestingly, if one leaves the scale parameter~$r_0$ free, this results into a factor in front of~$d\Omega^2$ in~(\ref{sol-1}) that cannot be removed by further redefinitions. This in turn implies a deficit of the solid angle. Consequently, the general solution~(\ref{alpha_beta_gamma}) describes a more general Bertotti-Robinson-like solution with a 3-cone.

Now we have to verify that the solution is realized for at list for a pair of real $\omega$ and $\lambda$. Setting, just as a first exemple, $\gamma_2 = - 1$ and $\lambda = 2$ and using (\ref{fe-2}), (\ref{A}), (\ref{gamma2}), we find that $\omega$ is a solution of the following equation:
\begin{equation}
47\omega^2 - 18\omega - 72 = 0,
\end{equation}
which admits real roots. In particular, the root $\omega = 1.444$ is plausible, because it leads to $A = 0.098$ implying $\alpha_0$ real if $R_0$ is negative. On the other hand, if we set $\lambda = 0$ we find $\omega=-3/2$. It is straightforward to check that these two values for $\lambda$ and $\omega$ can also provide $\alpha_0$ real if $R_0$ is negative. It is important to emphasise that the choice of $\lambda = 0$ is more motivated as it will be more clearly shown in the Section 4. The reason behind this choice is that for this specific value the PPN parameters of the Brans-Dicke-Rastall gravity coincide with those arising from GR, regardless of the value of $\omega$. Therefore, this particular value ensures the fulfillment of the local tests.

Finally, for completeness we remind the structure of the Robinson-Bertotti solution rediscovered also by Lovelock~\cite{lovelock-1,lovelock-2}. This space-time is a direct product~\cite{clement} of constant curvature spaces~$AdS_{2}\times S^{2}$. It possesses two apparent singularities~$r=0$ and $r=+\infty$. The former is associated to the origin of the coordinates and is likely unphysical, because curvature invariants stay finite there (this issue will be studied elsewhere), and the latter corresponds to a null hyper-surface. Indeed, by transformation
\begin{eqnarray}
\label{coorTR}
 r(\eta,\chi)&=&\frac{1}{\sin{(\chi-\eta)}}\,, \nonumber \\
 t(\eta,\chi)&=&\chi+\cot{(\chi-\eta)}\,.
\end{eqnarray}
the metric is reduced to the form:
\begin{equation}
\label{sol-1-ext-1}
 ds^{2}=d\eta^{2}-\cos^{2}{(\eta-\chi)}d\chi^{2}-d\Omega^{2}\,.
\end{equation}
The ``singularity'' $r=+\infty$ corresponds to~$(\chi-\eta)\rightarrow +0$, $\theta={\rm const}$, $\varphi=\rm const$. One can easily see that this hypersurface is null and regular (``horizon'').

\section{The PPN parameters}\label{PPN}

The fact that the only static, spherically symmetric exact solutions that can represent a star is the Schwarzschild one does not imply that necessarily the
classical tests of gravitational phenomena are automatically recovered. Instead, we must analyze the parametrized post-newtonian approach. In this section,
we will follow very closely the approach given in the reference \cite{weinberg} for the Brans-Dicke theory.

To analyze the PPN parameters is more convenient to write the equations as,
\begin{align}
R_{\mu\nu} &= \frac{8\pi}{\phi}\biggr\{T_{\mu\nu} + \frac{\lambda k_2}{2(1 - 2\lambda)}g_{\mu\nu}T\biggl\}  + \frac{\omega}{\phi^2}\biggr\{\phi_{;\mu}\phi_{\nu}
+ \frac{(\lambda - 1)}{2(1 - 2\lambda)}k_2g_{\mu\nu}\phi_{;\rho}\phi^{;\rho}\biggl\} + \frac{\phi_{;\mu;\nu}}{\phi}, \\
\Box\phi &= \frac{1}{k_1}\biggr\{8\pi\lambda T - \omega(1 - \lambda)\frac{\phi_{;\rho}\phi^{;\rho}}{\phi}\biggl\},
\end{align}
where
\begin{equation}
k_1 = 3\lambda - 2(1 - 2\lambda)\omega, \quad k_2 = 1 + \frac{\lambda - 2}{k_1}.
\end{equation}
When $\lambda = 1$, we come back to the usual Brans-Dicke equations,
\begin{eqnarray}
R_{\mu\nu} &=& \frac{8\pi}{\phi}\biggr\{T_{\mu\nu} + \frac{1 + \omega}{3 + 2\omega}g_{\mu\nu}T\biggl\}  + \frac{\omega}{\phi^2}\phi_{;\mu}\phi_{\nu}+\frac{\phi_{;\mu;\nu}}{\phi}, \\
\Box\phi &=& \frac{8\pi}{3 + 2\omega} T,
\end{eqnarray}
with
\begin{equation}
k_1 = 3 + 2\omega, \quad k_2 = 2\frac{(1 + \omega)}{3 + 2 \omega}.
\end{equation}
\par
We consider the following expansion for the metric using an expansion in the slow-motion, weak-field approximation~\cite{willb}:
\begin{eqnarray}
g_{00} &=& 1 + {\stackrel{2}{g}}_{00} + {\stackrel{4}{g}}_{00} + ...\\
g_{0i} &=& {\stackrel{3}{g}}_{0i} + ...\\
g_{ij} &=& - \delta_{ij} + {\stackrel{2}{g}}_{ij} + ...
\end{eqnarray}
For the Ricci tensor, the expansion takes the form,
\begin{eqnarray}
R_{00} &=& {\stackrel{2}{R}}_{00} + {\stackrel{4}{R}}_{00} + ...\\
R_{0i} &=& {\stackrel{3}{R}}_{0i} + ...\\
R_{ij} &=& {\stackrel{2}{R}}_{ij} + ...
\end{eqnarray}
while for the energy momentum tensor, we have,
\begin{eqnarray}
T^{00} = \stackrel{0}{T^{00}} + \stackrel{2}{T^{00}} + ...\quad &,& \quad T_{00} = \stackrel{0}{T^{00}} + \stackrel{2}{T^{00}} + 2{\stackrel{2}{g}}_{00}\stackrel{0}{T^{00}}...\\\
T^{0i} = \stackrel{1}{T_{0i} }+ ...\quad &,& \quad T_{0i} = - \stackrel{1}{T_{0i} }+ ...\\
T^{ij} = \stackrel{2}{T^{ij}} + ...\quad &,& \quad T_{ij} = \stackrel{2}{T^{ij}} + ...\\
T = \stackrel{0}{T^{00}} + \stackrel{2}{T^{00}} &+& \stackrel{2}{g_{00}}\stackrel{0}{T^{00}} - \stackrel{2}{T^{kk}} + ...
\end{eqnarray}
For the scalar field, we write,
\begin{equation}
\phi = \phi_0(1 + \xi) \equiv \frac{1}{G_0}(1 + \xi),
\end{equation}
with
\begin{equation}
\xi = {\stackrel{2}{\xi}} + {\stackrel{4}{\xi}} + ...
\end{equation}
\par
The expansion in the Einstein's equation reads:
\begin{eqnarray}
\stackrel{2}{R}_{00} &=& 8\pi G_0\biggr\{1 + \frac{\lambda k_2}{2(1 - 2\lambda)}\biggl\}\stackrel{0}{T^{00}},\\
\stackrel{4}{R}_{00} &=& 8\pi G_0\biggr\{\biggr[1 + \frac{\lambda k_2}{2(1 - 2\lambda)}\biggl]\stackrel{2}{T^{00}} + 2\biggr[1 + \frac{\lambda k_2}{2(1 - 2\lambda)}\biggl]{\stackrel{2}{g}}_{00}\stackrel{0}{T^{00}}\biggr.\nonumber\\
\biggr. &-& \stackrel{2}{\xi}\biggr(1 + \frac{\lambda k_2}{2(1 - 2\lambda)}\biggl)\stackrel{0}{T^{00}} - \frac{\lambda k_2}{2(1 - 2\lambda)}\stackrel{2}{T^{kk}}\biggl\}  \nonumber\\
&-& \omega\frac{(\lambda - 1)k_2}{2(1 - 2\lambda)}\stackrel{2}{\xi_{,k}}\stackrel{2}{\xi_{,k}} + \stackrel{2}{\xi_{,0,0}} - \stackrel{2}{\Gamma^k_{00}}\stackrel{2}{\xi_{,k}},\\
\stackrel{3}{R_{0i}} &=& - 8\pi G_0\stackrel{1}{T^{0i}} + \stackrel{2}{\xi_{,0,i}}, \\
\stackrel{2}{R_{ij}} &=&  - 8\pi G_0\frac{\lambda k_2}{2(1 - 2\lambda)}\delta_{ij}\stackrel{0}{T^{00}} + \stackrel{2}{\xi_{,i,j}},\\
\nabla^2\stackrel{2}{\xi} &=& - \frac{8\pi\lambda G_0}{k_1}\stackrel{0}{T^{00}}.
\end{eqnarray}
\par
The expansion of the Ricci's tensor up to second order, using the harmonic coordinate condition, gives:
\begin{eqnarray}
\stackrel{2}{R}_{00} &=& - \frac{1}{2}\nabla^2{\stackrel{2}{g}}_{00},\\
\stackrel{4}{R}_{00} &=& - \frac{1}{2}\biggr\{\nabla^2{\stackrel{4}{g}}_{00} - \partial^2_t{\stackrel{2}{g}}_{00} - {\stackrel{2}{g}}_{ij}\partial^2_{ij}{\stackrel{2}{g}}_{00} + \partial_k{\stackrel{2}{g}}_{00}\partial_k{\stackrel{2}{g}}_{00}\biggl\},\\
\stackrel{3}{R}_{0i} &=& - \frac{1}{2}\nabla^2{\stackrel{3}{g}}_{0i},\\
\stackrel{2}{R}_{ij} &=& - \frac{1}{2}\nabla^2{\stackrel{2}{g}}_{ij}.
\end{eqnarray}
Hence, the PPN equations are,
\begin{eqnarray}
\label{ppn1}
\nabla^2{\stackrel{2}{g}}_{00} &=& - 16\pi G_0\biggr\{1 + \frac{\lambda k_2}{2(1 - 2\lambda)}\biggl\}\stackrel{0}{T^{00}},\\
\nabla^2{\stackrel{4}{g}}_{00} &-& \partial^2_t{\stackrel{2}{g}}_{00} - {\stackrel{2}{g}}_{ij}\partial^2_{ij}{\stackrel{2}{g}}_{00} + \partial_k{\stackrel{2}{g}}_{00}\partial_k{\stackrel{2}{g}}_{00} = \nonumber\\
&-&16\pi G_0\biggr\{\biggr[1 + \frac{\lambda k_2}{2(1 - 2\lambda)}\biggl]\stackrel{2}{T^{00}} + 2\biggr[1 + \frac{\lambda k_2}{2(1 - 2\lambda)}\biggl]{\stackrel{2}{g}}_{00}\stackrel{0}{T^{00}}\biggr.\nonumber\\ \biggr.
&-& \stackrel{2}{\xi}\biggr(1 + \frac{\lambda k_2}{2(1 - 2\lambda)}\biggl)\stackrel{0}{T^{00}} - \frac{\lambda k_2}{2(1 - 2\lambda)}\stackrel{2}{T^{kk}}\biggl\}  \nonumber\\
&+& 2\omega\frac{(\lambda - 1)k_2}{2(1 - 2\lambda)}\stackrel{2}{\xi_{,k}}\stackrel{2}{\xi_{,k}} - 2\stackrel{2}{\xi_{,0,0}} + \partial_k\stackrel{2}{g_{00}}\stackrel{2}{\xi_{,k}},\\
\nabla^2{\stackrel{3}{g}}_{0i} &=& 16\pi G_0\stackrel{1}{T^{0i}} - 2\stackrel{2}{\xi_{,0,i}}, \\
\nabla^2\stackrel{2}{g}_{ij} &=& 16\pi G_0\frac{\lambda k_2}{2(1 - 2\lambda)}\delta_{ij}\stackrel{0}{T^{00}} - 2\stackrel{2}{\xi_{,i,j}},\\
\nabla^2\stackrel{2}{\xi} &=& - \frac{8\pi\lambda G_0}{k_1}\stackrel{0}{T^{00}}.
\end{eqnarray}
The Brans-Dicke limit ($\lambda = 1$) gives:
\begin{eqnarray}
\nabla^2{\stackrel{2}{g}}_{00} &=& 16\pi G_0\biggr\{\frac{2 + \omega}{3 + 2\omega}\biggl\}\stackrel{0}{T^{00}},\\
\label{ppn2}
\nabla^2{\stackrel{4}{g}}_{00} &-& \partial^2_t{\stackrel{2}{g}}_{00} - {\stackrel{2}{g}}_{ij}\partial^2_{ij}{\stackrel{2}{g}}_{00} + \partial_k{\stackrel{2}{g}}_{00}\partial_k{\stackrel{2}{g}}_{00} = \nonumber\\
& &16\pi G_0\biggr\{\biggr[\frac{2 + \omega}{3 + 2\omega}\biggl]\stackrel{2}{T^{00}} + 2\biggr[\frac{2 + \omega}{3 + 2\omega}\biggl]{\stackrel{2}{g}}_{00}\stackrel{0}{T^{00}}\biggr.\nonumber\\ \biggr.
&-& \stackrel{2}{\xi}\biggr(\frac{2 + \omega}{3 + 2\omega}\biggl)\stackrel{0}{T^{00}} + \frac{( 1 + \omega)}{3 + 2\omega}\stackrel{2}{T^{kk}}\biggl\}  \nonumber\\
&+& 2\stackrel{2}{\xi_{,0,0}} - \partial_k\stackrel{2}{g_{00}}\stackrel{2}{\xi_{,k}},\\
\nabla^2{\stackrel{3}{g}}_{0i} &=& - 16\pi G_0\stackrel{1}{T^{0i}} + 2\stackrel{2}{\xi_{,0,i}}, \\
\label{ppn4}
\nabla^2{\stackrel{3}{g}}_{ij} &=&  16\pi G_0\frac{1 + \omega}{3 + 2\omega}\delta_{ij}\stackrel{0}{T^{00}} + 2\stackrel{2}{\xi_{,i,j}},\\
\label{ppn5}
\nabla^2\stackrel{2}{\xi} &=& - \frac{8\pi G_0}{3 + 2\omega}\stackrel{0}{T^{00}}.
\end{eqnarray}
\par
In order to reproduce the Poisson's equation,
\begin{equation}
\nabla^2\Psi = 4\pi G\stackrel{0}{T^{00}},
\end{equation}
we write in equation (\ref{ppn1}), ${\stackrel{2}{g}}_{00} = - 2\Psi$, obtaining as for the gravitational coupling,
\begin{equation}
G = 2\biggr\{1 + \frac{\lambda k_2}{2(1 - 2\lambda)}\biggl\}G_0,
\end{equation}
which reduces to expression \eqref{efetivo} in the general case, and to the usual Brans-Dicke relation
\begin{equation}
G = \frac{4 + 2\omega}{3 + 2\omega}G_0,
\end{equation}
when $\lambda = 1$.
\par
The consistency of the previous result with (\ref{ppn5}) implies,
\begin{equation}
\stackrel{2}{\xi} =  - \biggr\{1 + \frac{\lambda k_2}{2(1 - 2\lambda)}\biggl\}^{-1}\frac{\lambda}{k_1}\Psi.
\end{equation}
Equation (\ref{ppn4}) may be written as,
\begin{equation}
\nabla^2\biggr\{\stackrel{2}{g}_{ij} - 2\biggr[1 + \frac{\lambda k_2}{2(1 - 2\lambda)}\biggl]^{-1}\frac{\lambda k_2}{2(1 - 2\lambda)}\delta_{ij}\Psi\biggl\} =
2\biggr[1 + \frac{\lambda k_2}{2(1 - 2\lambda)}\biggl]^{-1}\frac{\lambda}{k_1}\Psi_{,i,j}.
\end{equation}
Using $\Psi = - \frac{GM}{r}$ and following \cite{weinberg}, the solution reads,
\begin{equation}
\stackrel{2}{g}_{ij}  = - \frac{\lambda}{k_1{\cal A}}\biggr\{1 + \frac{k_2k_1}{1 - 2\lambda}\biggl\}\frac{GM}{r}\delta_{ij} +
\frac{\lambda}{k_1 {\cal A}}\biggr\{\frac{x_i x_j}{r^3}GM + 2GMR^2\biggr(\delta_{ij} - 3\frac{x_ix_j}{r^2}\biggl)\frac{1}{r^3}\biggl\},
\end{equation}
with the definitions,
\begin{equation}
GMR^2 = \int_0^\infty\biggr[\Psi + \frac{GM}{r}\biggl]r^2dr, \quad {\cal A} = 1 + \frac{\lambda k_2}{2(1 - 2\lambda)}.
\end{equation}
\par
To determine $\stackrel{4}g_{00}$, we use equation (\ref{ppn2}) in the vacuum, static case:
\begin{eqnarray}
\nabla^2{\stackrel{4}{g}}_{00} - {\stackrel{2}{g}}_{ij}\partial_i\partial_ j{\stackrel{2}{g}}_{00} + \partial_k{\stackrel{2}{g}}_{00}\partial_k{\stackrel{2}{g}}_{00} =\omega\frac{\lambda - 1}{1 - 2\lambda}\partial_k\stackrel{2}{\xi}\partial_k\stackrel{2}{\xi} + \partial_k\stackrel{2}g_{00}\stackrel{2}{\xi_{,k}}.
\end{eqnarray}
The solution is:
\begin{eqnarray}
{\stackrel{4}{g}}_{00} = - \frac{B}{2}\frac{G^2M^2}{r^2} - \frac{2\lambda}{{\cal A}k_1}\frac{G^2M^2R^2}{r^4} + \frac{c_1}{R}\frac{G^2 M^2}{r},
\end{eqnarray}
where $c_1$ is a new constant and
\begin{eqnarray}
B = 4 - \frac{2\lambda}{k_1 {\cal A}} - \omega(\lambda - 1)\frac{\lambda^2k_2}{(1 - 2\lambda)(k_1{\cal A})^2}.
\end{eqnarray}
Using the redefinitions described in \cite{weinberg}, we find finally:
\begin{eqnarray}
{\stackrel{2}{g}}_{00} &=& 2\frac{GM}{r},\\
{\stackrel{4}{g}}_{00} &=& - \frac{B}{2}\frac{G^2 M^2}{r^2} = - (\gamma - 1 + 2\beta)\frac{G^2 M^2}{r^2},\\
{\stackrel{2}{g}}_{ij}  &=& - \frac{\lambda}{k_1{\cal A}}\biggr\{1 + \frac{k_2k_1}{1 - 2\lambda}\biggl\}\frac{GM}{r}\delta_{ij} +
\frac{\lambda}{k_1 {\cal A}}\frac{x_i x_j}{r^3}GM\nonumber\\ &=&  (3\gamma - 1)\frac{GM}{r}\delta_{ij} +
(1 - \gamma)\frac{x_i x_j}{r^3}GM.
\end{eqnarray}
\par
The PPN parameters read,
\begin{eqnarray}
\gamma &=& 1 - \frac{\lambda}{k_1{\cal A}},\\
\beta &=& \frac{1}{2}\biggr\{\frac{B}{2} + \frac{\lambda}{k_1 {\cal A}}\biggl\} = 1 - \omega\frac{\lambda^2(\lambda - 1)k_2}{4(1 - 2\lambda)(k_1{\cal A})^2}.
\end{eqnarray}
The values $\gamma =1$ and $\beta = 1$ (General Relativity results) are possible if $\lambda = 0$. Even for $\lambda \neq 0$ the values of
$\gamma$ and $\beta$ can be very near those allowed by the observational constraints for a region in the $\lambda$ and $\omega$ parameter space.
Remark that, for $\lambda = 1$ the Brans-Dicke results,
\begin{eqnarray}
\gamma = \frac{1 + \omega}{2 + \omega}, \quad \beta = 1,
\end{eqnarray}
are recovered.
\par
Since the experimental tests give values for $\gamma$ and $\beta$ near 1 with a precision up to $10^{-5}$ \cite{willb}, we can consider that we must have $\lambda = 0$. However, a numerical inspection shows that values for the PPN parameters inside those constraints can be obtained in other regions of the parameter space of $\lambda$ and $\omega$. These regions include the usual Brans-Dicke case for which $\omega >> 1$.

\section{Cosmology}\label{cosmology}

Let us consider an isotropic and homogeneous space-time described by the flat Friedmann-Lema\^{\i}tre-Robertson-Walker (FLRW) metric,
\begin{equation}
ds^2 = dt^2 - a(t)^2(dx^2 + dy^2 + dz^2).
\end{equation}
In this case the equations of motion read:
\begin{eqnarray}
\label{em1}
\dot\rho + 3\frac{\dot a}{a}(1 + \textrm{w})\rho &=& - \frac{3(1 - \lambda)}{8\pi}\phi\biggr[\frac{\dddot a}{a} + \frac{\dot a}{a}\frac{\ddot a}{a} - 2\biggr(\frac{\dot a}{a}\biggl)^3\biggl],\\
\label{em2}
3\biggr(\frac{\dot a}{a}\biggl)^2 &=& \frac{8\pi \rho}{\phi}\biggr\{\frac{1 - 3\lambda}{2(1 - 2\lambda)} + \frac{3(1 - \lambda)}{2(1 - 2\lambda)}\textrm{w}\biggl\}\nonumber\\
&+& \omega\biggr[\frac{2 - 3\lambda}{2(1 - 2\lambda)}\biggl]\biggr(\frac{\dot\phi}{\phi}\biggl)^2
+ \biggr[\frac{3(1 - \lambda)}{2(1 - 2\lambda)}\frac{\ddot\phi}{\phi} + \frac{3(1 + \lambda)}{2(1 - 2\lambda)}\frac{\dot a}{a}\frac{\dot \phi}{\phi}\biggl],\\
\label{em3}
2\frac{\ddot a}{a} + \biggr(\frac{\dot a}{a}\biggl)^2 &=& - \frac{8\pi}{\phi}\biggr\{\frac{1 - \lambda - (1 + \lambda)\textrm{w}}{2(1 - 2\lambda)}\biggl\}\rho + \omega\frac{\lambda}{2(1 - 2\lambda)}\biggr(\frac{\dot\phi}{\phi}\biggl)^2\nonumber\\ &+& \frac{1 + \lambda}{2(1 - 2\lambda)}\frac{\ddot \phi}{\phi} + \frac{5 - \lambda}{2(1 - 2\lambda)}\frac{\dot a}{a}\frac{\dot\phi}{\phi},\\
\label{em4}
\frac{\ddot\phi}{\phi} + 3\frac{\dot a}{a}\frac{\dot\phi}{\phi} &=& \frac{8\pi\lambda}{3\lambda - 2(1 - 2\lambda)\omega}(1 - 3\textrm{w})\frac{\rho}{\phi}\nonumber\\
&-& \omega\frac{1 - \lambda}{3\lambda - 2(1 - 2\lambda)\omega}\biggr(\frac{\dot\phi}{\phi}\biggl)^2.
\end{eqnarray}
\par
Equations (\ref{em1})-(\ref{em4}) form a rich and complex system. In order to get a hint on which kind of solutions they predict, we consider power-law solutions, in the first place. The power-law solutions constitute a very restrictive case, but they can indicate the kind of cosmological solution we can expect from the BDR theory. Hence, suppose the solutions have the form:
\begin{equation}
\label{ansat}
a = a_0t^s, \quad \phi = \phi_0t^p, \quad \rho = \rho_0t^q,
\end{equation}
where $a_0$, $\phi_0$, $\rho_0$, $s$, $p$ and $q$ are constants.
\par
Plugging (\ref{ansat}) into (\ref{em1})-(\ref{em3}), we obtain the following relations:
\begin{align}
\label{r1}
\frac{\rho_0}{\phi_0}[q + 3 s(1 + \textrm{w})] &= \frac{3(\lambda - 1)}{4\pi}s(1 - 2s),\\
\label{r2}
3s^2 &= \frac{8\pi\rho_0}{\phi_0}\biggr[\frac{1 - 3\lambda + 3(1 - \lambda)\textrm{w}}{2(1 - 2\lambda)}\biggl] + \omega\frac{2 - 3\lambda}{2(1 - 2\lambda)}p^2\nonumber\\ &+
3\frac{(1 - \lambda)p(p - 1) + (1 + \lambda)sp}{2(1 - 2\lambda)},\\
\label{r3}
p[p - 1 + 3s] &= \frac{8\pi\lambda}{3\lambda - 2(1 - 2\lambda)\omega}\frac{\rho_0}{\phi_0}(1 - 3\textrm{w}) - \omega\frac{1 - \lambda}{3\lambda - 2(1 -2\lambda)\omega}p^2.
\end{align}
The condition to have consistent power-law solutions is $q = p - 2$. Using this relation and combining equations (\ref{r1})-(\ref{r3}), we obtain two coupled polynomials for $s$ and $p$:
\begin{eqnarray}
6s^2(1 - 2\lambda)[p - 2 + 3s(1 + \textrm{w})] =\nonumber\\ 6(\lambda - 1)s(1 - 2s)[1 - 3\lambda + 3(1 - \lambda)\textrm{w}] \nonumber\\
+ \{\omega p^2(2 - 3\lambda) + 3p[(1 - \lambda)(p - 1) + (1 + \lambda)s]\}[p - 2 + 3s(1 + \textrm{w})],\\
6[(\lambda - 1)s(1 - 2s)](1 - 3\textrm{w})\lambda =\nonumber\\ \{[p(p - 1) + 3sp][3\lambda - 2(1 - 2\lambda)\omega] + \omega(1 - \lambda)p^2\}[p - 2 + 3s(1 + \textrm{w})].
\end{eqnarray}
This system admits eight pairs of roots for $(s,p)$. For the dust case, $\textrm{w} = 0$, one of the pairs corresponds to the Minkowski case, $p = s = 0$. Another one is $s = p = 1/2$. A third root implies a curious configuration with $s = 0$ and $p = 2$, that is, a static universe, with a varying gravitational coupling. Among the other five pairs, two incorporate an accelerated regime of the expansion while remaining three describe a decelerating universe.
%\begin{figure*}[!t]
%\includegraphics[width=0.4\linewidth]{Fig1a.eps}\includegraphics[width=0.4\linewidth]{Fig1b.eps}\\
%\includegraphics[width=0.4\linewidth]{Fig1c.eps}\includegraphics[width=0.4\linewidth]{Fig1d.eps}
%\caption{The left panel represents the first root for $s$ displaying acceleration of the scale factor, the red line separating positive (left) and negative (right) values. The center left panel represents the first root for $p$, the red line separating positive (second and fourth quadrants) and negative values (first and third quadrants).
%In the center right and right panels, the value of $s$ (left), and $p$ (center) as function of $\lambda$ for
%$\omega = - 30$ are displayed.}
%\end{figure*}
\begin{center}
\begin{figure*}[!t]
\begin{minipage}[t]{0.22\linewidth}
\includegraphics[width=\linewidth]{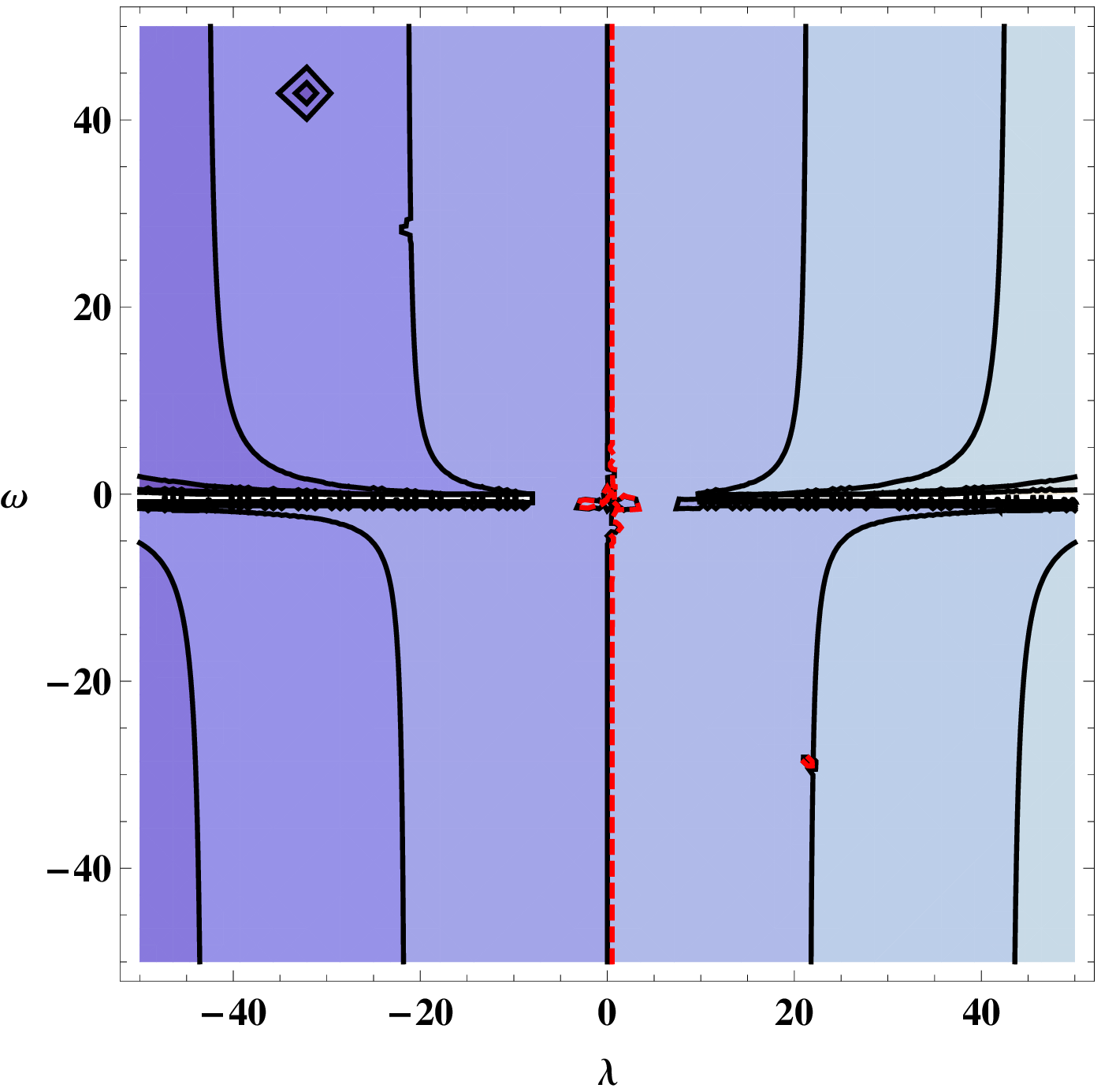}
\end{minipage} \hfill
\begin{minipage}[t]{0.22\linewidth}
\includegraphics[width=\linewidth]{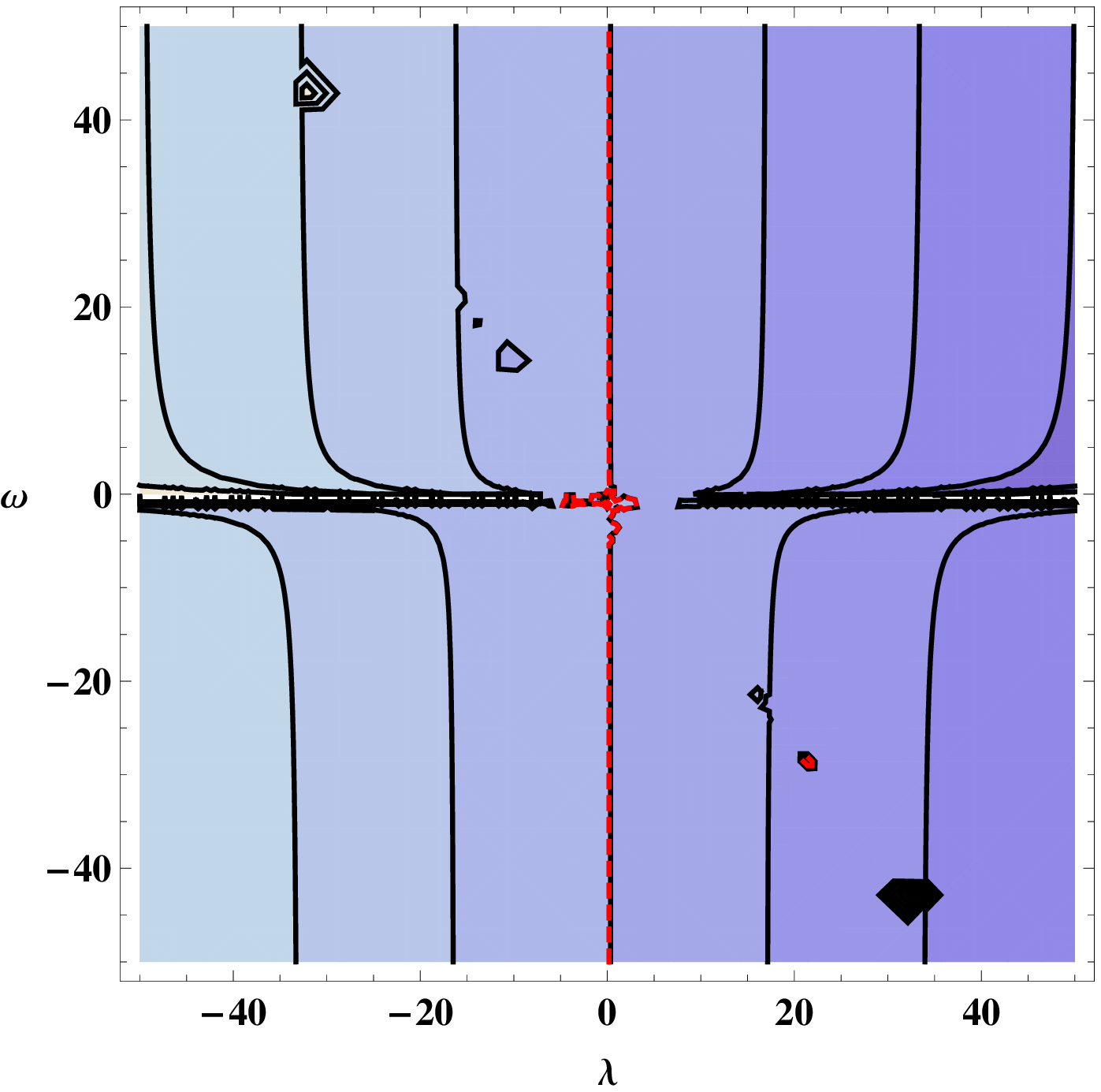}
\end{minipage} \hfill
\begin{minipage}[t]{0.22\linewidth}
\includegraphics[width=\linewidth]{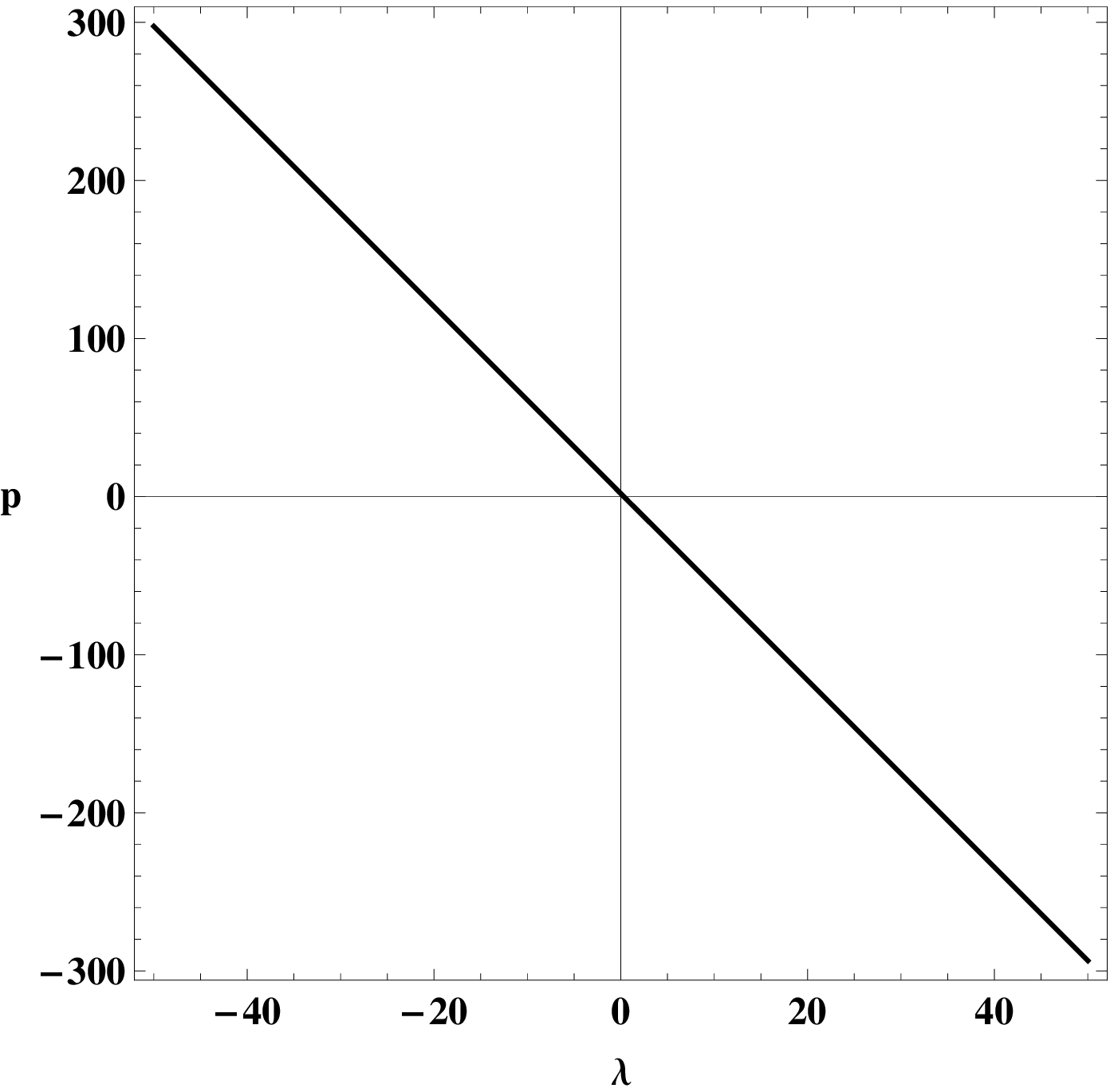}
\end{minipage} \hfill
\begin{minipage}[t]{0.22\linewidth}
\includegraphics[width=\linewidth]{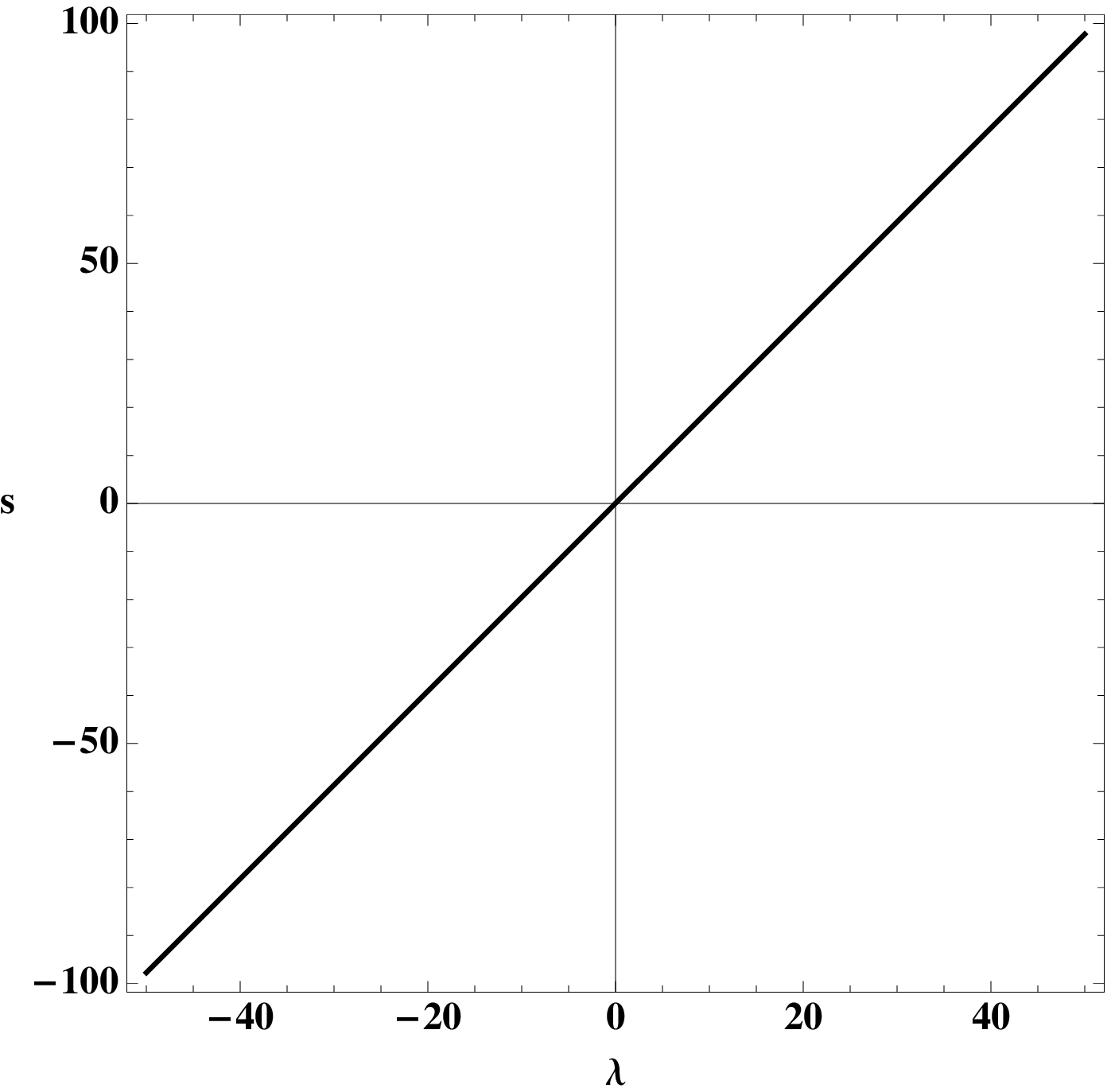}
\end{minipage} \hfill
\caption{The left panel represents the first root for $s$ displaying acceleration of the scale factor, the red line separating positive (left) and negative (right) values. The center left panel represents the first root for $p$, the red line separating positive (second and fourth quadrants) and negative values (first and third quadrants).
In the center right and right panels, the value of $s$ (right), and $p$ (center) as function of $\lambda$ for
$\omega = - 30$ are displayed.}
\end{figure*}
\end{center}
The set of equations (\ref{em1})-(\ref{em4}) may be recast into the form of a dynamical system:
\begin{align}
\dot H &= \frac{1}{2 k_3}\biggr\{- 6(1 - 2\lambda)(1 + \textrm{w})H^2 + [2 + (1 - \textrm{w})\omega]f^2 + 2\dot f + 2(2 + 3\textrm{w})Hf\biggl\},\\
\dot f &= \frac{1}{k_1 k_3 + 3\lambda(1 - 3\textrm{w})(1 - \lambda)}\biggr\{6\lambda(1 - 3\textrm{w})(1 - 2\lambda)H^2 \nonumber\\
&- \biggr[\omega[\lambda(1 - 3\textrm{w})(2 - 3\lambda) + (1 - \lambda)k_3] + k_1 k_3 + 3\lambda(1 - 3\textrm{w})(1 - \lambda)\biggr]f^2\biggr.\nonumber\\ \biggr. &- 3[k_1 k_3 + 3(1 + \lambda)\lambda(1 - 3\textrm{w})]Hf\biggr\},
\end{align}
where
\begin{equation}
k_{3} = 1 - 3\lambda + 3(1 - \lambda)\textrm{w}.
\end{equation}
This dynamical system is very complex, and depends not only on the values of $\lambda$ and $\omega$, but also on the value of the initial conditions for
$H$ and $f$. We look for an example of a deceleration/acceleration transition during the matter dominated phase ($\textrm{w} = 0$). Figure~\ref{Hubble-deceleration} displays as an example the behaviour of the Hubble function and deceleration parameter $q = - 1 - \frac{\dot H}{H^2}$ for $\omega = 1$ and $\lambda = - 1$, undergoing this transition. Note that the effective $G > 0$ stays positive~(see~(\ref{efetivo})).

\begin{center}
\begin{figure}[!t]
\begin{minipage}[t]{0.4\linewidth}
\includegraphics[width=\linewidth]{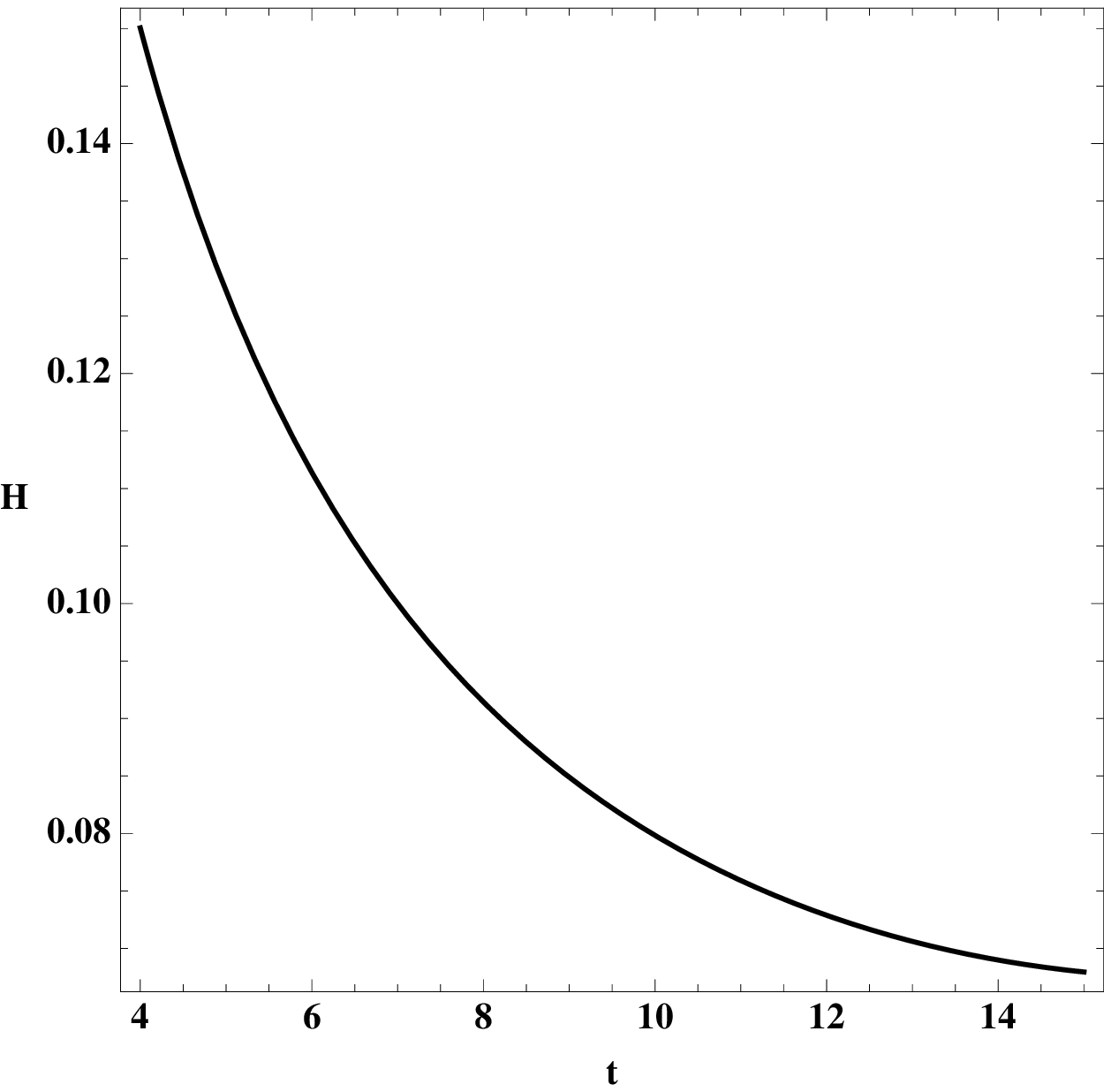}
\end{minipage} \hfill
\begin{minipage}[t]{0.4\linewidth}
\includegraphics[width=\linewidth]{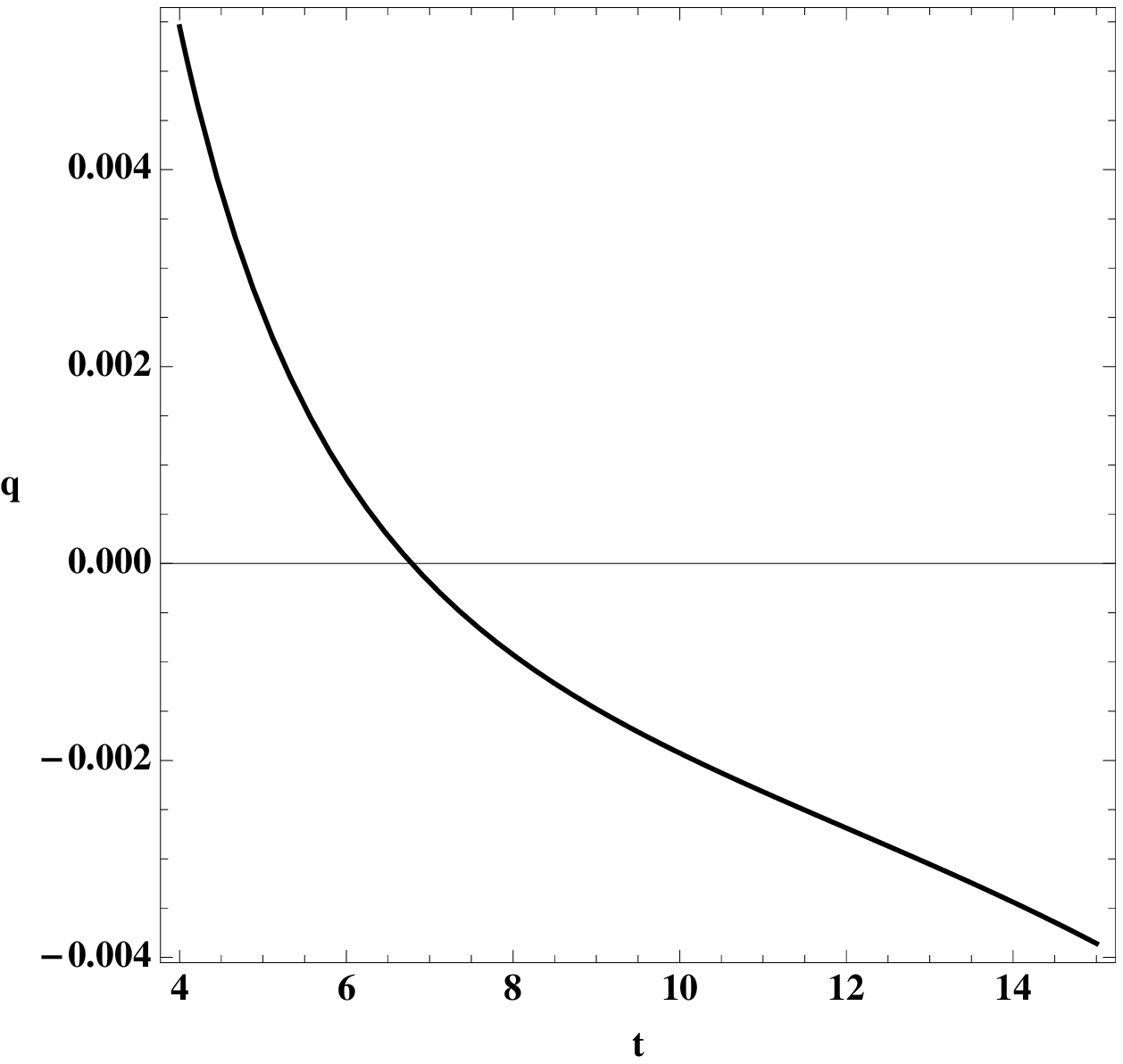}
\end{minipage} \hfill
\caption{Behaviour of the Hubble function $H$ and deceleration parameter $q$ for $\omega = 1$, $\lambda = - 1$ and $\textrm{w} = 0$.}\label{Hubble-deceleration}
\end{figure}
\end{center}

Consequently, in the framework of the BDR theory there are solutions that have a positive gravitational coupling and display a deceleration/acceleration transition in a matter dominated universe, thus, circumventing the restriction existing in the pure Brans-Dicke theory~\cite{f1}.
\par
However, as it was shown in the preceding section, the local tests are satisfied in the present theory at least for two cases: $\lambda = 0$, leading to the
PPN parameters identical to the GR ones; for nonvanishing $\lambda$ and $\omega$ very big. For the specific case $\lambda = 0$ and a matter dominate universe, the dynamical system reduces to,
\begin{eqnarray}
\label{ds1bis}
\dot H &=& -3H^2  + \biggr(\frac{\omega + 1}{2}\biggl)u^2 - uH,\\
\label{ds2bis}
\dot u &=& -3Hu - \frac{1}{2}u^2,
\end{eqnarray}
with the definition $u = \dot\phi/\phi$. The gravitational coupling is always positive in this case. There are power law accelerated solutions for $- \frac{7}{6} > \omega > - \frac{5}{4}$.
Even if our goal here is not to perform a complete dynamical system analysis, we display the trajectories of the solutions in the $(H,u)$ phase space for the
system (\ref{ds1bis},\ref{ds2bis}) in figure 3, where it is shown the region where the accelerated solutions occur indicate by the red lines. The origin corresponds to the Minkowski space-time. It can be seen that the accelerate/decelerate transition occurs, the universe reaching the decelerate phase in the future. Hence, the transition occurs in the opposite sense with respect to the previous example. This case seems more appropriate to describe a primordial phase.

\begin{center}
\begin{figure}[!t]
\includegraphics[width=\linewidth]{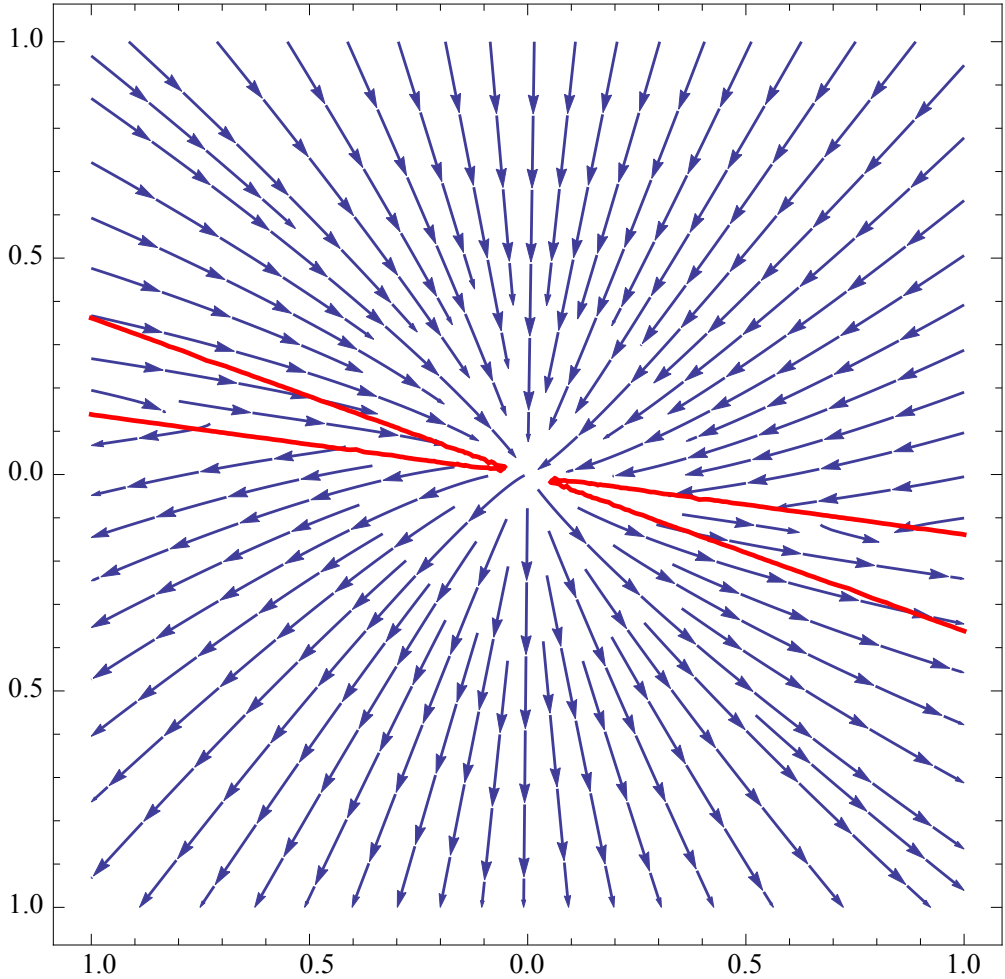}
\caption{Solutions in the $(H,\dot\phi/\phi) \equiv (H,u)$ phase space for $\lambda = 0$ and $\omega = - 1.2$. The vertical axis corresponds to $H$ and the horizontal one to $u$. The origin corresponds to Minkowski space-time. The red lines delimite the region where
the expansion of the universe is accelerated.}
\end{figure}
\end{center}
\par

For the radiation phase, the ordinary solutions of the standard model are also present here, since for this case $T=g^{\mu\nu}T_{\mu\nu} = 0$.  This is important in order not to spoil the success of the standard cosmological model, mainly in what concerns the primordial nucleosynthesis.

\section{Conclusions}\label{conclusions}

In this work we have combined the idea of a scalar-tensor theory of the Brans-Dicke type and Rastall's proposal of a gravitational anomaly encoded in the violation of the conventional conservation law for the energy-momentum tensor. In doing so, we end up with two free parameters: the usual Brans-Dicke parameter
$\omega$ and Rastall's parameter $\lambda$, representing a degree of the non-conservation. The resulting theory referred to as the BDR (Brans-Dicke-Rastall) theory cannot
be derived from an action principle, at least in the Riemannian context, as it usually happens the Rastall-type theory, and as we can expect from a theory that tries to classically incorporate effects
typical of the quantum regime. But, there are claims that an action principle can be recovered using a more general geometrical framework -- for example, the Weyl geometry \cite{smalleybis,romero}.
\par
We have investigated the BDR theory in two contexts: spherically symmetric static solutions and cosmological regime. In the first case, we found that the only possible non-trivial analytical solution is a Robinson-Bertotti type solution, which represents a kind of {\it stretched star}. In the General Relativity context such a
solution emerges from a configuration including electromagnetic field while in the BDR theory it is a vacuum solution. The only possible solution in the BDR theory
that can represent a star is the usual Schwarzschild solution corresponding to the trivial configuration where the scalar field is constant. Based on this result we can
argue that the usual classical tests of General Relativity are equally satisfied in the BDR theory without any important restriction on the parameters $\omega$ and
$\lambda$. The case $\lambda = 0$ looks like a scalar-tensor generalization of the case studied in \cite{Gao:2009me}, which is a particular case of the usual Rastall's theory, with very particular and interesting properties.
\par
For the cosmological case, we found power law solutions for the matter dominated phase, some of them representing an accelerating expansion, others, decelerating. This fact suggests that perhaps a decelerating/accelerating transition can be achieved in the matter dominated phase in the BDR theory. In fact, we find some particular solutions where this transition occurs, in one sense or in another, leading to possible models for the present or the primordial universe. But the results in general are not only very sensitive to the parameters $\omega$ and $\lambda$, but also to the initial conditions. In this sense, a more detailed dynamical analysis must be performed, which we hope to present in the future.

\section*{Acknowledgements}
We thank FAPES (Brazil) and CNPq (Brazil) for partial financial support. TRPC and VS are grateful to CNPq for support.

\end{document}